\documentclass[aps,pra,preprint,superscriptaddress,showpacs,showkeys]{revtex4}
\usepackage{amssymb,bm}
\usepackage{graphicx}
\usepackage{amsmath}
\usepackage{epstopdf}
\usepackage{float}
\allowdisplaybreaks

\begin{document}

\title{Charge asymmetry in the differential cross section of high-energy
 bremsstrahlung  in the  field of a heavy atom}

\author{P.A. Krachkov}\email{peter_phys@mail.ru}
\affiliation{Budker Institute of Nuclear Physics, 630090 Novosibirsk, Russia}
\affiliation{Novosibirsk State University, 630090 Novosibirsk, Russia}
\author{A. I. Milstein}\email{A.I.Milstein@inp.nsk.su}
\affiliation{Budker Institute of Nuclear Physics, 630090 Novosibirsk, Russia}

\date{\today}

\begin{abstract}
The distinction between  the charged particle and antiparticle  differential cross sections of  high-energy  bremsstrahlung  in the electric  field of a heavy atom is investigated. The consideration is based on the quasiclassical approximation to the wave functions in the external field. The charge asymmetry (the ratio of the antisymmetric and symmetric parts of the differential cross section) arises due to the account for the first  quasiclassical correction to the differential cross section. All evaluations  are performed  with the exact account of the atomic field. We consider in detail the charge asymmetry for electrons and muons. For electrons, the nuclear size effect is not important while for muons this effect should be taken into account. For the longitudinal polarization of the initial charged particle, the account for the first quasiclassical correction to the differential cross section leads to the  asymmetry in the cross section with respect to the replacement $\varphi\rightarrow-\varphi$, where $\varphi$ is the azimuth angle between the photon momentum and the momentum of the final charged particle.
\end{abstract}



\maketitle

\section{Introduction}
The theoretical investigation of  high-energy bremsstrahlung and high-energy particle-antiparticle photoproduction  in the electric field of a heavy nucleus or  atom  has a long history because of importance of these  processes for various applications; for the latter process see reviews in Refs. \cite{HO1980, H2000}. These processes should be taken into account when considering electromagnetic showers in detectors, they also give   the significant part of the radiative corrections in many cases. Therefore, it is necessary to know the cross sections of these processes with high accuracy.
In the Born approximation, the   cross sections of both processes   have been obtained for arbitrary  energies of particles  and for arbitrary atomic form factors \cite{BH1934, Racah1934} (see also  Ref.~\cite{BLP1982}). The Coulomb corrections to the cross section, which are the difference between the exact in the parameter  $\eta=Z\alpha$ cross section and the Born cross section, are  very important (here  $Z$ is the atomic charge number, $\alpha$ is the fine-structure constant, $\hbar=c=1$). There are  formal expressions for the Coulomb corrections to the cross sections exact in $\eta$ and energies of particles \cite{Overbo1968}. However, the numerical computations based on these expressions  become more and more difficult when energies are increasing, and, for instance,  the numerical results for $e^+e^-$ photoproduction have been obtained so far only for the photon energy  $\omega<12.5\,$~MeV  \cite{SudSharma2006}.

At high energies of initial particles, the final particle momenta usually have small angles with respect to the incident direction. In this case  typical angular momenta, which provide  the main contribution to the cross section, are large ($l\sim E/\Delta\gg 1$, where $E$ is energy and  $\Delta$ is the momentum transfer).  This is why the quasiclassical approximation, based on the account of  large angular momenta  contributions,  becomes applicable.   In this approximation, the wave functions and the Green's functions of the Dirac equation in the external field have very simple forms which drastically simplify their  use in specific calculations. The wave functions, obtained in the leading quasiclassical approximation for the Coulomb field, are the famous Furry-Sommerfeld-Maue wave functions \cite{Fu, ZM} (see also Ref.~\cite{BLP1982}).  The quasiclassical Green's function  have been derived in Ref.~ \cite{MS1983} for the case of a pure Coulomb field, in Ref.~\cite{LM95A} for an arbitrary spherically symmetric field, in Ref.~\cite{LMS00} for a localized field which generally possesses no spherical symmetry, and in Ref.~\cite{DM2012} for combined strong laser and atomic fields.

In the leading  quasiclassical approximation, the cross sections  for   pair photoproduction and bremsstrahlung have been obtained in  \cite{BM1954,DBM1954,OlsenMW1957,O1955,OM1959}. The first quasiclassical corrections to the spectra of both processes, as well as to the total cross section of pair photoproduction, have been obtained in Refs. \cite{LMS2004,LMSS2005,DM10,DM12}.  Recently, the  first quasiclassical correction to the fully differential cross section was obtained in Ref.~\cite{LMS2012} for $e^+e^-$  pair photoproduction and in Ref.~\cite{DLMR2014} for $\mu^+\mu^-$ pair  photoproduction. As a result, the charge asymmetry in these processes (the asymmetry of the cross section with respect to permutation of particle and antiparticle momenta)  was predicted. This asymmetry is absent in the cross section calculated in the Born approximation and also in the cross section exact in the parameter $\eta$ but calculated in the leading quasiclassical approximation. Thus, the charge asymmetry appears solely  due to  the quasiclassical corrections to the Coulomb corrections. The difference between the atomic field and the Coulomb field of a nucleus results in the modification of the cross sections (effect os screening). The influence of  screening on the Coulomb corrections to  $e^+e^-$  pair photoproduction is small  for the differential cross section and for the total cross section \cite{DBM1954}. However, screening is important  for the Born term. The quantitative investigation of the effect of screening on  the Coulomb corrections to the  photoproduction cross section is performed in Ref.~\cite{LMS2004}.

The influence   of screening on the  bremsstrahlung cross section in an atomic field is more complicated.  It is shown in Refs.~\cite{OlsenMW1957,LMSS2005} that the Coulomb corrections to the differential cross section are very susceptible to screening. However, the Coulomb corrections to the cross section integrated over the momentum of final charged particle (electron or muon)
are independent of screening in the leading approximation over a small parameter $1/m_er_{scr}$, where $r_{scr}\sim Z^{-1/3}(m_e\alpha)^{-1}$ is a screening
radius and  $m_e$ is the electron mass. The quantitative investigation of the effect of screening on  the Coulomb corrections to the  spectrum of bremsstrahlung is performed in Ref.~\cite{LMSS2005}.
The differential cross section of  bremsstrahlung, calculated in the leading quasiclassical approximation, is the same for $e^+$ and $e^-$ (for $\mu^+$ and  $\mu^-$). Therefore,  to  predict the charge asymmetry (the difference between the  bremsstrahlung differential cross section for particles  and antiparticles in the atomic field), one should  perform  calculations in the next-to-leading quasiclassical approximation. This is the main goal of our paper. The result is obtained  exactly in the parameter $\eta$. Besides, for the case of muons the nuclear size effect is taken into account.

The bremsstrahlung differential cross section from high-energy  charged particle in an atomic field, $d\sigma(\bm p,\bm q,\bm k,\eta)$, can be written as
 \begin{eqnarray}\label{sigsa}
&&d\sigma(\bm p,\bm q,\bm k,\eta)=d\sigma_s(\bm p,\bm q,\bm k,\eta)+d\sigma_a(\bm p,\bm q,\bm k,\eta)\,,\nonumber\\
&&d\sigma_s(\bm p,\bm q,\bm k,\eta)=\frac{d\sigma(\bm p,\bm q,\bm k,\eta)+d\sigma(\bm p,\bm q,\bm k,-\eta)}{2}\,,\nonumber\\
&& d\sigma_a(\bm p,\bm q,\bm k,\eta)=\frac{d\sigma(\bm p,\bm q,\bm k,\eta)-d\sigma(\bm p,\bm q,\bm k,-\eta)}{2}\,,
\end{eqnarray}
where $\bm k$ is the photon momentum, $\bm p$ and $\bm q$ are the initial and final charged particle momenta, respectively. Evidently, the bremsstrahlung differential cross section from high-energy  antiparticle  can be obtained from $d\sigma(\bm p,\bm q,\bm k,\eta)$ by the replacement $\eta\rightarrow -\eta$, so that it is equal to   $d\sigma_s(\bm p,\bm q,\bm k,\eta)-d\sigma_a(\bm p,\bm q,\bm k,\eta)$. We show that the antisymmetric part of the differential  cross section, $d\sigma_a(\bm p,\bm q,\bm k,\eta)$, is independent of screening in the kinematical region which provides the main contribution to the antisymmetric part  of the spectrum.

The paper is organized as follows. In Sec.~\ref{general} we derive the general expression for the quasiclassical matrix element of the process. In Sec.~\ref{functions} we find in the quasiclassical approximation all structures of the Green's function of the squared Dirac equation for a charged particle  in arbitrary localized potential and the corresponding structures of the wave functions. We obtain the leading terms and the first quasiclassical corrections as well. Using these wave functions, we derive in Sec.~\ref{matrixelement} the matrix element of the process and the corresponding differential cross section for arbitrary localized potential and in the particular case of the pure Coulomb field. In Sec.~\ref{CAE} we investigate in detail the charge asymmetry in  high-energy  bremsstrahlung from electrons. In this case the nuclear size effect is not important. In Sec.~\ref{CAM} we investigate  the charge asymmetry in  high-energy  bremsstrahlung from muons which is sensitive to the deviation at small distances of the nuclear atomic field from the pure Coulomb field. Finally, in Sec.~\ref{concl} the main conclusions of the paper are presented.

\section{General discussion}\label{general}
The differential cross section of  bremsstrahlung  in the electric  field of a heavy atom  reads~\cite{BLP1982}
\begin{equation}\label{eq:cs}
d\sigma=\frac{\alpha\omega q\varepsilon_{ q}}{(2\pi)^4}\,d\Omega_{\bm{k}}\,d\Omega_{\bm{q}}\,d\omega|M|^{2}\,,
\end{equation}
where $d\Omega_{\bm{k}}$ and $d\Omega_{\bm{q}}$ are  the solid angles corresponding to the photon momentum $\bm k$ and the final charged particle momentum $\bm q$,
$\omega=\varepsilon_{ p}-\varepsilon_{ q}$ is the photon energy, $\varepsilon_{ p}=\sqrt{ \bm{p}^2+m^2}$,  $\varepsilon_{ q}=\sqrt{ \bm q^2+m^2}$, and $m$ is the particle mass. Below we assume that $\varepsilon_{ p}\gg m$ and $\varepsilon_{ q}\gg m$.   The matrix element $M$ reads
\begin{equation}\label{M12}
M=\int d\bm r \,\bar u_{\bm q }^{(-)}(\bm r )\,\bm\gamma\cdot
\bm e^*\,u _{\bm p}^{(+)}(\bm r )\exp{(-i\bm k\cdot\bm r )}\,\,,
\end{equation}
where  $\gamma^\mu$ are the Dirac matrices, $ u_{\bm p}^{(+)}(\bm r )$  and $u_{\bm q}^{(-)}(\bm r )$ are the solutions of the Dirac equation in the external field, $\bm e$ is the photon polarization vector. The superscripts $(-)$ and $(+)$ remind us that the asymptotic forms of $ u_{\bm q}^{(-)}(\bm r )$ and $ u_{\bm p}^{(+)}(\bm r )$ at large $\bm r$ contain, in addition to the plane wave, the spherical convergent and divergent waves, respectively. The wave functions $ u_{\bm p}^{(+)}(\bm r )$  and $u_{\bm q}^{(-)}(\bm r )$ have the form \cite{DLMR2014}
\begin{eqnarray}\label{wfD1}
&&\bar u_{\bm q }^{(-)}(\bm r )=\bar u_{\bm q }[f_0(\bm r,\bm q)-\bm\alpha\cdot\bm f_1(\bm r,\bm q)-\bm\Sigma\cdot\bm f_2(\bm r,\bm q)]\,,\nonumber\\
&&u _{\bm p}^{(+)}(\bm r )=[g_0(\bm r,\bm p)-\bm\alpha\cdot\bm g_1(\bm r,\bm p)-\bm\Sigma\cdot\bm g_2(\bm r,\bm p)]u _{\bm p}\,,\nonumber\\
&& u_{\bm p}=\sqrt{\frac{\varepsilon_p+m}{2\varepsilon_p}}
\begin{pmatrix}
\phi\\
\dfrac{{\bm \sigma}\cdot {\bm
p}}{\varepsilon_p+m}\phi
\end{pmatrix}\,,\quad
u_{\bm q}=\sqrt{\frac{\varepsilon_q+m}{2\varepsilon_q}}
 \begin{pmatrix}
\chi\\
\dfrac{{\bm \sigma}\cdot {\bm
q}}{\varepsilon_q+m}\chi
\end{pmatrix}\,,
\end{eqnarray}
where $\phi$ and  $\chi$  are spinors, $\bm\alpha=\gamma^0\bm\gamma$, $\bm\Sigma=\gamma^0\gamma^5\bm\gamma$, and $\bm\sigma$ are the Pauli matrices. The following relations hold
\begin{eqnarray}\label{relfg}
g_0(\bm r,\bm q)=f_0(\bm r,-\bm q)\,,\quad \bm g_1(\bm r,\bm q)=\bm f_1(\bm r,-\bm q)\,,\quad \bm g_2(\bm r,\bm q)=-\bm f_2(\bm r,-\bm q)\,.
\end{eqnarray}
The wave functions in the atomic field  can be found from the Green's function $D(\bm r_2,\,\bm r_1|\varepsilon)$ of the  ``squared'' Dirac equation in this
field using the relations \cite{DLMR2014}
 \begin{eqnarray}\label{Green1}
&&\frac{\exp{ipr_1}}{4\pi r_1}u_{{\bm p}}^{(+)}({\bm r}_2)=-
\lim_{ r_1\to \infty }D({\bm r}_2,{\bm r}_1|\varepsilon_{ p}){u}_{{\bm p}}\,,\quad \bm p=-p\bm n_1\,
,\nonumber\\
&&\frac{\exp{iqr_2}}{4\pi r_2}\bar{u}_{{\bm q}}^{(-)}({\bm r}_1)=-
\lim_{ r_2\to \infty }\bar{u}_{{\bm q}}D({\bm r}_2,{\bm r}_1|\varepsilon_{ q})\,,\quad \bm q=q\bm n_2\,,\nonumber\\
\end{eqnarray}
where $\bm n_1=\bm r_1/r_1$, $\bm n_2=\bm r_2/r_2$, and
\begin{eqnarray}\label{FGD}
&&D(\bm r_2,\,\bm r_1|\varepsilon)=\langle \bm r_2|\frac{1}{\hat{\cal P}^2-m^2+i0}| \bm r_1\rangle\nonumber\\
&&=\langle \bm r_2|\left[(\varepsilon-V(r))^2+\bm\nabla^2-m^2+i\bm\alpha\cdot\bm\nabla V(r) +i0\right]^{-1}| \bm r_1\rangle\,.
\end{eqnarray}
Here $\hat{\cal P}=\gamma^\mu{\cal P}_\mu$, ${\cal P}_\mu=(\varepsilon-V(r),i\bm\nabla)$, and $V(r)$ is the atomic potential. It follows from Eq.~\eqref{FGD}
that the Green's function $D(\bm r_2,\,\bm r_1|\varepsilon)$ can be written as
\begin{eqnarray}\label{Dd}
&&D(\bm r_2,\,\bm r_1|\varepsilon)=d_0(\bm r_2,\bm r_1)+\bm\alpha\cdot\bm d_1(\bm r_2,\bm r_1)+\bm\Sigma\cdot\bm d_2(\bm r_2,\bm r_1)\,.
\end{eqnarray}

It is convenient to calculate the matrix element for definite helicities of the particles. Let $\mu_p$ , $\mu_q$, and $\lambda$ be the signs of the helicities of  initial charged particle, final  charged particle, and photon, respectively.   We fix the coordinate system so that $\bm\nu=\bm k/\omega$  is directed along $z$-axis and $\bm q$ lies in the $xz$ plane with $q_{x}>0$. Denoting helicities by the subscripts, we have
\begin{align}\label{spinors}
\phi_{\mu_p}&=\frac{1+\mu_p\bm \sigma\cdot\bm n_p}{4\cos(\theta_{p}/2)}
\begin{pmatrix}1+\mu_p\\1-\mu_p\end{pmatrix}
\approx \frac14\left(1+\frac{\theta_{p}^2}{8}\right)\left(1+\mu_p\bm \sigma\cdot\bm n_p\right)\begin{pmatrix}1+\mu_p\\1-\mu_p\end{pmatrix}\,,
\nonumber\\
\chi_{\mu_q}&=\frac{1+\mu_q\bm \sigma\cdot\bm n_q}{4\cos(\theta_{q}/2)}
\begin{pmatrix}1+\mu_q\\1-\mu_q\end{pmatrix}
\approx \frac14\left(1+\frac{\theta_{q}^2}{8}\right)\left(1+\mu_q\bm \sigma\cdot\bm n_q\right)\begin{pmatrix}1+\mu_q\\1-\mu_q\end{pmatrix}\,,
\nonumber\\
\bm e_{\lambda}&=\frac{1}{\sqrt{2}}(\bm e_x+i\lambda\bm e_y)\,,
\end{align}
where $\theta_{p}$ and $\theta_{q}$ are the polar angles of the vectors $\bm p$ and $\bm q$, respectively. The unit vectors $\bm e_x$ and $\bm e_y$ are directed along $\bm q_{\perp}$
 and $\bm k \times \bm q$, respectively, where the notation $\bm X_\perp=\bm X -(\bm X\cdot\bm \nu)\bm \nu$ for any vector $\bm X$ is used.  We also  introduce the vectors $\bm \theta_{p}=\bm p_{\perp}/p$, $\bm \theta_{q}=\bm q_{\perp}/q$, and the matrix  ${\cal F}=u_{{\bm p}\,\mu_p}\bar{u}_{{\bm q}\,\mu_q}$, which can be written as
\begin{eqnarray}\label{calf}
&&{\cal F}=\frac{1}{8}(a_{\mu_p\mu_q}+\bm\Sigma\cdot \bm b_{\mu_p\mu_q})[\gamma^0(1+PQ)+\gamma^0\gamma^5(P+Q)+(1-PQ)-\gamma^5(P-Q)],\nonumber\\
&&P=\frac{\mu_pp}{\varepsilon_p+m}\,,\quad Q=\frac{\mu_qq}{\varepsilon_q+m}\,,
\end{eqnarray}
where $a_{\mu_p\mu_q}$ and $\bm b_{\mu_p\mu_q}$ are defined from
\begin{equation}
\phi_{\mu_p}\chi_{\mu_q}^\dagger=\frac{1}{2}(a_{\mu_p\mu_q}+\bm\sigma\cdot \bm b_{\mu_p\mu_q})\,.
\end{equation}
We obtain from Eq.\eqref{spinors}
\begin{eqnarray}\label{ab}
&&a_{\mu\mu}=1-\frac{\theta_{pq}^2}{8}-\frac{i\mu}{4}\bm\nu\cdot[\bm\theta_p\times\bm\theta_q]\,,\nonumber\\
&&  a_{\mu\bar\mu}=\frac{\mu}{\sqrt{2}}\bm e_\mu\cdot\bm\theta_{pq}\,,\nonumber\\
&&\bm b_{\mu\mu}=\left\{\mu\left[1-\frac{1}{8}(\bm\theta_{p}+\bm\theta_{q})^2\right]+\frac{i}{4}\bm\nu\cdot[\bm\theta_p\times\bm\theta_q]\right\}\bm\nu\nonumber\\ &&+\frac{\mu}{2}(\bm\theta_{p}+\bm\theta_{q})+\frac{i}{2}[\bm\theta_{pq}\times\bm\nu]\,,\nonumber\\
&&\bm b_{\mu\bar\mu}=\sqrt{2}\bm e_\mu-\frac{1}{\sqrt{2}}(\bm e_\mu,\bm\theta_{p}+\bm\theta_{q})\bm\nu\,,
\end{eqnarray}
where $\bm\theta_{pq}=\bm\theta_{p}-\bm\theta_{q}$ and $\bar\mu=-\mu$. The matrix element $M$, Eq.~\eqref{M12}, can be written as follows
\begin{multline}\label{MF}
M=\int d\bm r\, e^{-i\bm k\cdot\bm r}\, \mathrm{Sp}\{(f_0-\bm\alpha\cdot\bm f_1-\bm\Sigma\cdot\bm f_2)\bm\gamma\cdot\bm e_\lambda^*(g_0-\bm\alpha\cdot\bm g_1-\bm\Sigma\cdot\bm g_2){\cal F}\}
\,.
\end{multline}
Note that only the terms with  $(P+Q)$ and $(1+PQ)$ in $\cal F$, Eq.~\eqref{calf}, contribute to the matrix element (\ref{MF}) because it contains the odd  number of the gamma-matrices.

In the quasiclassical approximation the relative magnitude of the functions $f_0$, $\bm f_{1,2}$, $g_0$, and   $\bm g_{1,2}$ is different, so that
\begin{equation}
f_0\sim l_c f_1\sim l^2_c f_2\,,\quad
g_0\sim l_c g_1\sim l^2_c g_2\,,\quad
d_0\sim l_c d_1\sim l^2_c d_2\,,
\end{equation}
where $l_c\sim \varepsilon/\Delta\gg 1$ is the characteristic value of the angular momentum in the process, $\bm\Delta=\bm q+\bm k-\bm p$ is the momentum transfer. To find the distinction between the differential cross section of bremsstrahlung from particles  and antiparticles, it is necessary to take into account the first quasiclassical corrections to the functions $f_0$, $g_0$,  $\bm f_{1}$, and $\bm g_{1}$, while the functions  $\bm f_{2}$ and $\bm g_{2}$ can be taken in the leading quasiclassical approximation.
Let us introduce the quantities
\begin{eqnarray}\label{fg}
&&(A_{00},\,\bm A_{01},\,\bm A_{10},\,\bm A_{02},\,\bm A_{20})=\int \!\!d\bm r\,\exp{(-i\bm k\cdot\bm r )}(f_0g_0,\, f_0\bm g_1,\, \bm f_1 g_0,\,f_0\bm g_2\,, \bm f_2 g_0)\,.
\end{eqnarray}
In terms of these quantities, the matrix element $M$ has the form
\begin{eqnarray}\label{Mtot1}
&&M=\delta_{\mu_p\mu_q}\Big[\delta_{\lambda\mu_p}(\bm e^*_\lambda, -\bm\theta_{q}A_{00}-2\bm A_{10}+2\mu_p\bm A_{20})\nonumber\\
&&+\delta_{\lambda\bar\mu_p}(\bm e^*_\lambda, -\bm\theta_{p}A_{00}+2\bm A_{01}+2\mu_p\bm A_{02})\Big]-\frac{m\mu_p(p-q)}{\sqrt{2}\,pq}\delta_{\mu_q\bar\mu_p}\delta_{\lambda\mu_p}A_{00}\,.
\end{eqnarray}
Below we calculate all quantities in \eqref{Mtot1} for arbitrary atomic potential $V(r)$ which includes the effect of screening and the nuclear size effect as well.

\section{Green's functions and wave functions}\label{functions}
Let us consider the case of arbitrary central localized potential $V(r)$. We expand the Green's function $D(\bm r_2,\,\bm r_1|\varepsilon)$, Eq.~\eqref{FGD}, up to the second order with respect to the correction $\bm\alpha\cdot\bm\nabla V(r)$:
\begin{eqnarray}\label{FGDnew}
&&D(\bm r_2,\,\bm r_1|\varepsilon)=\langle \bm r_2|\frac{1}{{\cal H}}-\frac{1}{{\cal H}}i\bm\alpha\cdot\bm\nabla V(r)\frac{1}{{\cal H}}+
\frac{1}{{\cal H}}i\bm\alpha\cdot\bm\nabla V(r)\frac{1}{{\cal H}}i\bm\alpha\cdot\bm\nabla V(r)\frac{1}{{\cal H}}| \bm r_1\rangle\,,\nonumber\\
&&{\cal H}=\varepsilon^2-m^2-2\varepsilon\varphi(r)+\bm\nabla^2+i0\,,\quad \varphi(r)=V(r)-\frac{V^2(r)}{2\varepsilon}\,.
\end{eqnarray}
The function $D^{(0)}(\bm r_2,\,\bm r_1|\varepsilon)=\langle \bm r_2|{\cal H}^{-1}|\bm r_1\rangle$ is the Green's function of the Klein-Gordon equation. This function was found in the quasiclassical approximation with the first  correction taken into account \cite{LMS00}~:
\begin{eqnarray}\label{D0}
&& D^{(0)}(\bm r_2,\bm r_1 |\,\varepsilon )=
\frac{ie^{i\kappa r}}{4\pi^2r} \int d\bm Q \exp\left[iQ^2-i
r\int_0^1dx V(\bm R_x) \right]\nonumber\\
&&\times\left\{ 1+\frac{i r^3}{2\kappa}
\int\limits_0^1 dx \int\limits_0^x dy (x-y) \bm\nabla_\perp
V(\bm R_x)\cdot\bm\nabla_\perp  V(\bm R_y)\right\}\ ,\nonumber\\
&& \bm r=\bm r_2-\bm r_1\,,\quad \bm R_x= \bm r_1+x\bm r+\bm Q \sqrt{\frac{2r_1r_2}{\kappa r}}\,,
\end{eqnarray}
where $\bm Q$ is a two-dimensional vector perpendicular to $\bm r$  and $\bm\nabla_\perp$ is the component of the gradient perpendicular to $\bm r$.
Within the same accuracy,  $D^{(0)}(\bm r_2,\,\bm r_1|\varepsilon)$ coincides with the contribution  $d(\bm r_2,\,\bm r_1)$ to the Green's
 function $D(\bm r_2,\,\bm r_1|\varepsilon)$, Eq. (\ref{Dd}).

 Using this formula and Eqs.~\eqref{wfD1}, \eqref{Green1}, and \eqref{Dd},  we obtain  the function $f_0(\bm r,\bm q)$,
\begin{eqnarray}\label{f0}
&& f_0(\bm r,\bm q)=-\frac{i}{\pi}e^{-i\bm q\cdot\bm r}\int d\bm Q \exp\left[iQ^2-i
\int_0^\infty dx V(\bm r_x) \right]\nonumber\\
&&\times\left\{ 1+\frac{i}{2\varepsilon_q}
\int\limits_0^\infty dx \int\limits_0^x dy (x-y) \bm\nabla_\perp V(\bm r_x)\cdot\bm\nabla_\perp V(\bm r_y)\right\}\ ,\nonumber\\
&&\bm r_x= \bm r+x\bm n_{\bm q}+\bm Q \sqrt{\frac{2r}{\varepsilon_q}}\,, \quad \bm Q\cdot\bm n_{\bm q}=0\,,
\end{eqnarray}
where $\bm\nabla_\perp$ is the component of the gradient perpendicular to $\bm n_{\bm q}=\bm q/q$. Then we use the relation
\begin{equation}
i\bm\nabla V(r)=\frac{1}{2\varepsilon}[\bm p\,,\,{\cal H}]+\frac{i}{2\varepsilon}\bm\nabla V^2(r)\,,
\end{equation}
and write the linear in $\bm\nabla V(r)$ term in Eq. (\ref{FGDnew})  as  $\bm\alpha\cdot\bm d_1(\bm r_2,\bm r_1)$, where
\begin{eqnarray}\label{D1}
&&\bm d_1(\bm r_2,\bm r_1)=-\frac{i}{2\varepsilon}(\bm \nabla_1+\bm \nabla_2)D^{(0)}(\bm r_2,\bm r_1 |\,\varepsilon )+\delta\bm d_1(\bm r_2,\bm r_1)\,,\nonumber\\
&&\delta\bm d_1(\bm r_2,\bm r_1)=-\langle \bm r_2|\frac{1}{{\cal H}}\frac{i}{2\varepsilon}\bm\nabla V^2(r)\frac{1}{{\cal H}}| \bm r_1\rangle\,.
\end{eqnarray}
If we replace  $V(r)$ by $V(r)+\delta V(r)$ in the operator ${\cal H}$, where $\delta V(r)=-i\bm\alpha\cdot\bm\nabla V^2(r)/(2\varepsilon)^2$, then we obtain from Eq.~\eqref{D0}
\begin{eqnarray}\label{D11}
&&\delta\bm d_1(\bm r_2,\bm r_1)=-\frac{ie^{i\kappa r}}{16\pi^2\varepsilon^2} \int d\bm Q \exp\left[iQ^2-ir\int_0^1dx V(\bm R_x) \right]
\int\limits_0^1 dx \,\bm\nabla V^2(\bm R_x)\,,
\end{eqnarray}
where $\bm R_x$ is given in \eqref{D0}. Using  Eqs.~\eqref{wfD1}, \eqref{Green1}, and \eqref{Dd}, we find the function $\bm f_1(\bm r,\bm q)$,
\begin{eqnarray}\label{f1}
&& \bm f_1(\bm r,\bm q)=\frac{1}{2\varepsilon}(i\bm\nabla-\bm q)f_0(\bm r,\bm q)+\delta \bm f_1(\bm r,\bm q)\,,\nonumber\\
&&\delta \bm f_1(\bm r,\bm q)=-\frac{i}{4\pi\varepsilon^2}e^{-i\bm q\cdot\bm r}\int d\bm Q \exp\left[iQ^2-i\int_0^\infty dx V(\bm r_x) \right]
\int\limits_0^\infty dx \bm\nabla V^2(\bm r_x)\,.
\end{eqnarray}
where $\bm r_x$ is given in Eq.~\eqref{f0}.

To transform the third term in \eqref{FGDnew}, we replace $i\bm\nabla V(r)$ by $\frac{1}{2\varepsilon}[\bm p\,,\,{\cal H}]$. Then it follows from Eqs.~\eqref{wfD1}, \eqref{Green1}, and \eqref{Dd}  that the function $\bm d_2(\bm r_2,\bm r_1)$ is
\begin{eqnarray}\label{d2}
&&\bm d_2(\bm r_2,\bm r_1)=-\frac{i}{(2\varepsilon)^2}[\bm \nabla_2\times\bm \nabla_1]D^{(0)}(\bm r_2,\bm r_1 |\,\varepsilon )+\delta\bm d_2(\bm r_2,\bm r_1)\,,\nonumber\\
&&\delta\bm d_2(\bm r_2,\bm r_1)=\bm l_{21}\,\langle \bm r_2|\frac{1}{{\cal H}}\frac{ V'(r)}{2\varepsilon r}\frac{1}{{\cal H}}| \bm r_1\rangle\,,
\end{eqnarray}
where $\bm l_{21}=-(i/2)(\bm r_2\times\bm \nabla_2-\bm r_1\times\bm \nabla_1)$  and $V'(r)=\partial V(r)/\partial r$. In \eqref{d2}  we use the relation $[\bm l,{\cal H}]=0$.
If we replace  $V(r)$ by $V(r)+\delta V(r)$ in the operator ${\cal H}$, where $\delta V(r)=r^{-1}V'(r)/(2\varepsilon)^2$, then we obtain from Eq.~\eqref{D0}
\begin{eqnarray}\label{d22}
&&\delta\bm d_2(\bm r_2,\bm r_1)=\bm l_{21}\,\frac{e^{i\kappa r}}{16\pi^2\varepsilon^2} \int d\bm Q \exp\left[iQ^2-ir\int_0^1dx V(\bm R_x) \right]
\, \int\limits_0^1 dx \, \frac{V'(\bm R_x)}{R_x}\,.
\end{eqnarray}
Substituting this expression in \eqref{d2}, we finally find  $\bm d_2(\bm r_2,\bm r_1)$,
\begin{eqnarray}\label{d2tot}
&&\bm d_2(\bm r_2,\bm r_1)=-\frac{re^{i\kappa r}}{16\pi^2\varepsilon^2} \int d\bm Q \exp\left[iQ^2-ir\int_0^1dx V(\bm R_x) \right]
\nonumber\\
&&\times  \int\limits_0^1 dx\, \int\limits_0^x dy \,[\bm \nabla V(\bm R_x)\times \bm \nabla V(\bm R_y)]\,.
\end{eqnarray}
The corresponding  function $\bm f_2(\bm r,\bm q)$ is
\begin{eqnarray}\label{f2tot}
&&\bm f_2(\bm r,\bm q)=-\frac{e^{-i\bm q\cdot\bm r}}{4\pi \varepsilon^2} \int d\bm Q \exp\left[iQ^2-ir\int_0^1dx V(\bm r_x) \right]
\nonumber\\
&&\times  \int\limits_0^\infty dx\, \int\limits_0^x dy \,[\bm \nabla V(\bm r_x)\times \bm \nabla V(\bm r_y)]\,.
\end{eqnarray}

For the Coulomb field $V_c(r)=-\eta/r$, we find from \eqref{f0}, \eqref{f1}, and \eqref{f2tot}
 \begin{eqnarray}\label{f0f1f2}
&&f_0(\bm r,\bm q)= F_A+(1+\bm n_{\bm q}\cdot\bm n)F_C\,,\nonumber\\
&&\bm f_1(\bm r,\bm q)=(\bm n_{\bm q}+\bm n)\eta F_B\,,\nonumber\\
&&\bm f_2(\bm r,\bm q)=-i\bm\Sigma\cdot[\bm n_{\bm q}\times\bm n]F_C\,,
 \end{eqnarray}
where
\begin{eqnarray}\label{fun123QC}
&&F_A(\bm r,\,\bm q,\,\eta)=\exp\left(\frac{\pi\eta}{2}-i\bm q\cdot\bm r\right)
[\Gamma(1-i\eta)F(i\eta,1,\,iz)\nonumber\\
&&+\frac{\pi\eta^2{\mbox e}^{i\frac{\pi}{4}}}{2\sqrt{2qr}}\Gamma(1/2-i\eta)F(1/2+i\eta,1,\,iz)]\,,\nonumber\\
&&F_B(\bm r,\,\bm q,\,\eta)=-\frac{i}{2}\exp\left(\frac{\pi\eta}{2}-i\bm q\cdot\bm r\right)
[\Gamma(1-i\eta)F(1+i\eta,\,2,\,iz)\nonumber\\
&&+\frac{\pi\eta^2{\mbox e}^{i\frac{\pi}{4}}}{2\sqrt{2qr}}\Gamma(1/2-i\eta)F(3/2+i\eta,\,2,\,iz)]\,,\nonumber\\
&&F_C(\bm r,\,\bm q,\,\eta)=-\exp\left(\frac{\pi\eta}{2}-i\bm q\cdot\bm r\right)
\frac{\pi\eta^2{\mbox e}^{i\frac{\pi}{4}}}{8\sqrt{2qr}}\Gamma(1/2-i\eta)F(3/2+i\eta,2,\,iz)\,,\nonumber\\
&&z=(1+\bm n\cdot\bm n_{\bm q})qr\,,\quad \bm n=\frac{\bm r}{r}\,.
\end{eqnarray}
Here  $\Gamma(x)$ is the Euler Gamma function and  $F(\alpha,\beta,x)$   is the confluent hypergeometric  function.
The results \eqref{f0f1f2} and \eqref{fun123QC} are in agreement with that obtained in \cite{LMS2012}.

\section{Calculation of the matrix element}\label{matrixelement}

The calculation of the quantities $A_{00}$, $\bm A_{01}$, $\bm A_{10}$, $\bm A_{02}$, and $\bm A_{20}$ \eqref{fg} is performed in the same way as  in Ref.\cite{LMSS2005}.  We present  details of this very tricky calculation in Appendix.
We obtain
\begin{eqnarray}\label{ALLA}
&&A_{00}= \frac{1}{\omega m^4}\int d\bm r \exp\left[-i\bm\Delta\cdot\bm r-i\chi(\rho)\right]\Big[i2\varepsilon_p\varepsilon_q\xi_p\xi_q(\bm p_\perp+\bm q_\perp)
\nonumber\\
&&+m^2(\varepsilon_p\xi_p-\varepsilon_q \xi_q)\int\limits_0^\infty dx\,x \,\bm \nabla_\perp V(\bm r-x\bm\nu)\Big]\cdot\bm \nabla_\perp V(\bm r)\,,\nonumber\\
&&\bm A_{01}= \frac{\varepsilon_q\xi_q}{\omega m^2}\int d\bm r \exp\left[-i\bm\Delta\cdot\bm r-i\chi(\rho)\right]
\nonumber\\
&&\times\Big[i\bm \nabla_\perp V(\bm r)+\frac{\bm \Delta}{2\varepsilon_p}\int\limits_0^\infty dx\,x \,\bm \nabla_\perp V(\bm r-x\bm\nu)\cdot\bm \nabla_\perp V(\bm r)
+\frac{i}{2\varepsilon_p}\bm \nabla_\perp V^2(\bm r)\Big]\,,\nonumber\\
&&\bm A_{02}= -\frac{\varepsilon_q\xi_q}{2\omega\varepsilon_p m^2}\int d\bm r \exp\left[-i\bm\Delta\cdot\bm r-i\chi(\rho)\right]
\nonumber\\
&&\times\int\limits_0^\infty dx\, [\bm \nabla V(\bm r-x\bm\nu)\times\bm \nabla V(\bm r)]\,,\nonumber\\
&&\bm A_{10}=-\bm A_{01}(\varepsilon_q \leftrightarrow \varepsilon_p ,\,\xi_q \leftrightarrow \xi_p)\,,\quad
\bm A_{20}=-\bm A_{02}(\varepsilon_q \leftrightarrow \varepsilon_p ,\,\xi_q \leftrightarrow \xi_p)\,,\nonumber\\
&&\chi(\rho)=\int_{-\infty}^\infty V(z,\bm\rho)dz\,,\quad  \xi_p=\frac{m^2}{m^2+p_\perp^2}\,,\quad \xi_q=\frac{m^2}{m^2+q_\perp^2}\,.
\end{eqnarray}

 Substituting Eq.~\eqref{ALLA} in Eq.~\eqref{Mtot1}, we find  the matrix element $M$,
\begin{eqnarray}\label{Mres}
&&M=-\delta_{\mu_p\mu_q}(\varepsilon_p\delta_{\lambda\mu_p}+ \varepsilon_q\delta_{\lambda\bar\mu_p})
[N_0(\bm e^*_\lambda,\xi_p\bm p_\perp-\xi_q\bm q_\perp)+N_1(\bm e^*_\lambda,\varepsilon_p\xi_p\bm p_\perp-\varepsilon_q\xi_q\bm q_\perp)]\,  \nonumber\\
&&-\frac{1}{\sqrt{2}}m\mu_p\delta_{\mu_p\bar\mu_q}\delta_{\lambda\mu_p}(\varepsilon_p-\varepsilon_q)[N_0(\xi_p-\xi_q)+N_1(\varepsilon_p\xi_p-\varepsilon_q\xi_q)]\,,\nonumber\\
&&N_0=\frac{2i}{\omega m^2\Delta^2}\int d\bm r \exp\left[-i\bm\Delta\cdot\bm r-i\chi(\rho)\right]\bm\Delta\cdot\bm\nabla_\perp V(\bm r)\,,\nonumber\\
&&N_1=\frac{1}{\omega m^2\varepsilon_p\varepsilon_q}\int d\bm r \exp\left[-i\bm\Delta\cdot\bm r-i\chi(\rho)\right]\int\limits_0^\infty dx\,x \,\bm \nabla_\perp V(\bm r-x\bm\nu)\cdot\bm \nabla_\perp V(\bm r)\,.
\end{eqnarray}
Note that in Eq.~\eqref{Mtot1} the contributions of $\bm A_{02}$ and $\bm A_{20}$ cancel out the contributions of the terms with $\bm \nabla_\perp V^2(\bm r)$ in $\bm A_{01}$ and
 $\bm A_{10}$ \eqref{ALLA}. The amplitude $M$ is exact in the potential $V(r)$. It contains the leading quasiclassical contribution and the first quasiclassical correction as well. For high-energy
 bremsstrahlung from  electrons in the  field of a heavy atom, it is necessary to take into account the effect of screening. For high-energy
 bremsstrahlung from muons it is necessary to take also into account the finite nuclear radius $R$ (nuclear size effect), because the muon Compton wavelength, $\lambda_\mu=1/m_\mu=1.87\,\mbox{fm}$, is smaller than $R$, $R=7.3\,\mbox{fm}$ for gold and $R=7.2\,\mbox{fm}$ for lead,  $m_\mu$ is the muon mass.

 From \eqref{Mres} we have:
\begin{eqnarray}\label{M2}
&&\sum_{\lambda\, \mu_q}|M|^2=S_0+S_1+S_2\,,\nonumber\\
&&S_0=\frac{m^2|N_0|^2}{2}\left[\frac{\Delta^2}{m^2}(\varepsilon_p^2+\varepsilon_q^2)\xi_p\xi_q-2\varepsilon_p\varepsilon_q(\xi_p-\xi_q)^2\right]\,,\nonumber\\
&&S_1=\frac{m^2\mbox{Re}N_0N_1^*}{2}\,\Bigg\{\frac{\Delta^2}{m^2}(\varepsilon_p^2+\varepsilon_q^2)(\varepsilon_p+\varepsilon_q)\xi_p\xi_q\nonumber\\
&&+\Big[(\varepsilon_p^2+\varepsilon_q^2)(\varepsilon_p-\varepsilon_q)-4\varepsilon_p\varepsilon_q(\varepsilon_p\xi_p-\varepsilon_q\xi_q)\Big](\xi_p-\xi_q)\Bigg\}\,,\nonumber\\
&&S_2=-\mu_p\mbox{Im}N_0N_1^*\,\omega^2(\varepsilon_p+\varepsilon_q)\xi_p\xi_q\,[\bm p_\perp\times\bm q_\perp]\cdot\bm\nu\,.
\end{eqnarray}
 The quantity $S_0$ is the even  function of $\eta$, it contributes to the symmetric  term $d\sigma_s(\bm p,\bm q,\bm k,\eta)$ of the cross section  \eqref{sigsa}. The quantity $S_1$ is the odd  function of $\eta$, it contributes to the antisymmetric  term  $d\sigma_a(\bm p,\bm q,\bm k,\eta)$ of the cross section \eqref{sigsa}. The quantity $S_2$ is the even  function of $\eta$, it contributes to the symmetric  term $d\sigma_s(\bm p,\bm q,\bm k,\eta)$ of the cross section \eqref{sigsa} which vanishes after averaging over the helicity $\mu_p$ of the initial electron. Note that the contribution of $S_2$ to the cross section is responsible for the effect of asymmetry with respect to the replacement $\varphi_i\rightarrow -\varphi_i$, where
  $\varphi_i$ are the azimuth angles of the final particles in the frame where the z axis is directed along $\bm p$. Such asymmetry is absent in the cross section calculated in the leading quasiclassical approximation.  We emphasize that the  contributions $S_1$ and $S_2$  are nonzero due to accounting for the next-to-leading quasiclassical terms.

  The coefficients $N_0$ and $N_1$ depend on the momenta $\bm p$, $\bm q$, and  $\bm k$ via the momentum transfer $\bm\Delta$. Therefore, it is easy to find from \eqref{eq:cs} and \eqref{M2} the cross section  $d\sigma/d\omega d\bm\Delta_\perp$. A simple integration gives
\begin{eqnarray}\label{sigdelta}
&&\frac{d\sigma_s}{d\omega d\bm\Delta_\perp}=
\frac{\alpha\omega\varepsilon_q m^4}{(2\pi)^3\varepsilon_p}|N_0|^2\Phi(\zeta)\,,\quad
\frac{d\sigma_a}{d\omega d\bm\Delta_\perp}=
\frac{\alpha\omega\varepsilon_q (\varepsilon_p+\varepsilon_q) m^4}{(2\pi)^3\varepsilon_p}\mbox{Re}N_0N_1^*\Phi\,,\nonumber\\
&&\Phi= \frac{\ln(\zeta+\sqrt{1+\zeta^2})}{\zeta\sqrt{1+\zeta^2}}\left(\zeta^2  \frac{\varepsilon_p^2+\varepsilon_q^2}{\varepsilon_p\varepsilon_q}+1\right)-1\,,\quad
\zeta=\frac{\Delta_\perp}{2m}\,.
\end{eqnarray}
The function $\Phi$ has the following asymptotic forms:
\begin{eqnarray}\label{phi}
&&\Phi=\left(\frac{\varepsilon_p^2+\varepsilon_q^2}{\varepsilon_p\varepsilon_q}-\frac{2}{3}\right) \zeta^2\quad \mbox{at}\,\zeta\ll 1\,,\nonumber\\
&&\Phi=\frac{\varepsilon_p^2+\varepsilon_q^2}{\varepsilon_p\varepsilon_q}\ln(2\zeta)-1 \quad \mbox{at}\,\zeta\gg 1\,.
\end{eqnarray}

 \section{Charge asymmetry in  high-energy  bremsstrahlung from electrons}\label{CAE}
\begin{figure}[h]
\centering
\includegraphics[width=0.6\linewidth]{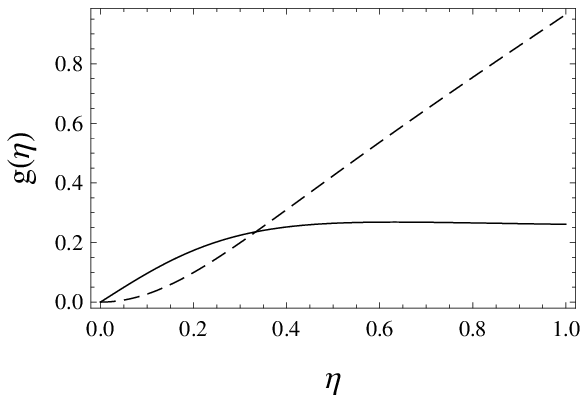}
\setlength{\unitlength}{0.7\linewidth}
\caption{The functions  $\mbox{Re}g(\eta)$ (solid curve) and  $\mbox{Im}g(\eta)$ (dashed curve), Eq.\eqref{M2}.}
\label{fig:g}
\end{figure}

  As is known, the main contribution to the Coulomb corrections to the symmetric part of the differential cross section of  bremsstrahlung is given by the region $\Delta\sim\mbox{max}(r_{scr}^{-1},\Delta_{min})$ \cite{OlsenMW1957,LMSS2005}, where $\Delta_{min}=p-q-\omega\approx m^2\omega/2\varepsilon_q\varepsilon_p$.
However, the main contribution to the charge asymmetry  is given by the region $\Delta\gg \mbox{max}(r_{scr}^{-1},\Delta_{min})$.
In this region we can neglect the effect of screening, replace  $V(r)$ by the Coulomb potential $V_c(r)=-\eta/r$,
and neglect also $\Delta_\parallel$ in comparison with $\Delta_\perp$. A simple calculation gives for the coefficient $N_0$ and $N_1$ in \eqref{Mres}:
\begin{eqnarray}\label{N0N1C}
&&N_0=\frac{8\pi\eta(L\Delta)^{2i\eta}}{\omega m^2\Delta^2}\frac{\Gamma(1-i\eta)}{\Gamma(1+i\eta)}\,,\quad
N_1=\frac{2\pi^2\eta^2(L\Delta)^{2i\eta}}{\omega m^2\varepsilon_p\varepsilon_q\Delta}\frac{\Gamma(1/2-i\eta)}{\Gamma(1/2+i\eta)}\,,
\end{eqnarray}
where $L\sim \mbox{min}(\varepsilon_p/m^2,\, r_{scr})$. Note that the factor $(L\Delta)^{2i\eta}$ is irrelevant because it disappears in $|M|^2$.
Then we  obtain for the coefficients in $S_0$, $S_1$, and $S_2$ \eqref{M2}:
\begin{eqnarray}\label{M2C}
&&|N_0|^2=\left(\frac{8\pi\eta}{\omega m^2\Delta^2}\right)^2\,,\quad \mbox{Re}N_0N_1^*=\frac{\pi\mbox{Re}g(\eta)\Delta}{4\varepsilon_p\varepsilon_q}|N_0|^2\,,\nonumber\\
&&\mbox{Im}N_0N_1^*=\frac{\pi\mbox{Im}g(\eta)\Delta}{4\varepsilon_p\varepsilon_q}|N_0|^2\,,\quad
g(\eta)=\eta\frac{\Gamma(1-i\eta)\Gamma(1/2+i\eta)}{\Gamma(1+i\eta)\Gamma(1/2-i\eta)}\,.
\end{eqnarray}
As it should, in the region $\Delta\sim m$ there are no  Coulomb corrections to $d\sigma_s(\bm p,\bm q,\bm k,\eta)$ calculated in the leading quasiclassical approximation.  The Coulomb corrections  to $d\sigma_a(\bm p,\bm q,\bm k,\eta)$ (the term $S_1$) and  $d\sigma_s(\bm p,\bm q,\bm k,\eta)$ (the term $S_2$) are accumulated in the functions $\mbox{Re} g(\eta)$ and  $\mbox{Im} g(\eta)$, respectively.
These functions are shown in Fig.\ref{fig:g}. At $\eta\ll 1$ we have $\mbox{Re} g(\eta)\approx \eta$ and $\mbox{Im} g(\eta)\approx -(4\ln 2)\eta^2$. It is seen from  Fig.\ref{fig:g} that the functions  $\mbox{Re} g(\eta)$ and $\mbox{Im} g(\eta)$ differ significantly from their small-$\eta$ asymptotic forms already at very small $\eta$.

 At $\omega\ll \varepsilon_p$, the ratio  of the antisymmetric part of the cross section to the symmetric one,
 \begin{eqnarray}\label{s1s01}
\frac{S_1}{S_0}=\frac{\pi\Delta\mbox{Re}g(\eta)}{2\varepsilon_p}\,,
\end{eqnarray}
increases with  $\Delta/\varepsilon_p$ and   can be more than ten percent. The ration $S_2/S_0$ is small at $\omega\ll\varepsilon_p$ because it is suppressed by the factor $(\omega/\varepsilon_p)^2$.

 If $\bm p_\perp\gg m$ and $\bm q_\perp\gg m$, then
   \begin{eqnarray}\label{s1s02}
\frac{S_1}{S_0}=\frac{\pi\mbox{Re}g(\eta)}{2\Delta}\bm\Delta\cdot\bm\theta_{qp}\,,
\quad \frac{S_2}{S_0}=\mu_p\frac{\pi\omega(\varepsilon_p+\varepsilon_q)\mbox{Im}g(\eta)}{2(\varepsilon_p^2+\varepsilon_q^2)\Delta}[\bm\Delta\times\bm\theta_{qp}]\cdot\bm\nu\,,
\end{eqnarray}
where $\bm\theta_{qp}=\bm p_\perp/p-\bm q_\perp/q$. Thus, the azimuth asymmetry increases with $\omega$ and may be important.

\begin{figure}[h]
\centering
\includegraphics[width=0.6\linewidth]{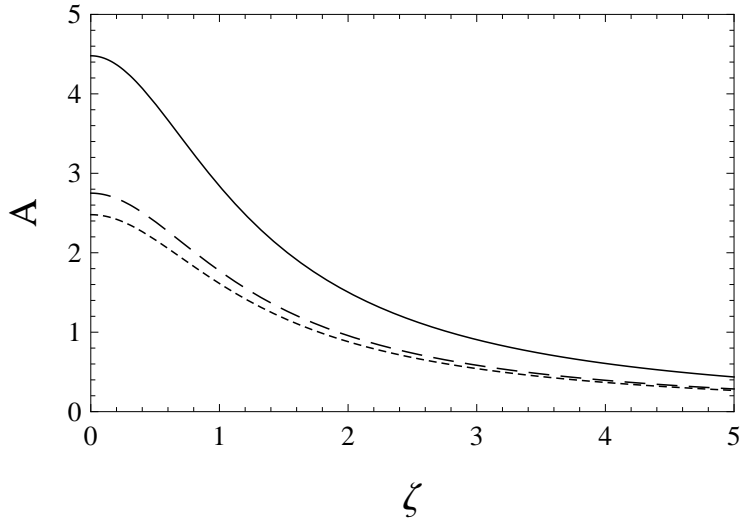}
\setlength{\unitlength}{0.7\linewidth}
\caption{Dependence of $A=\sigma_0^{-1}d\sigma_a/d\omega d\bm\Delta_\perp$ on $\zeta=\Delta_\perp/2m$, Eq.~\eqref{sigdelta1}, for a few values of $t=\varepsilon_q/\varepsilon_p$;
$\sigma_0=\alpha\eta^2\mbox{Re}g(\eta)/(2m^2\omega\varepsilon_p\Delta_\perp)$: $t=0.25$ (solid curve), $t=0.5$ (dashed curve), and $t=0.75$ (dotted curve).}
\label{fig:siga}
\end{figure}

Let us discuss the cross section $d\sigma/d\omega d\bm\Delta_\perp$ at $\Delta\gg \mbox{max}(r_{scr}^{-1},\Delta_{min})$, Eq.~\eqref{sigdelta}:
\begin{eqnarray}\label{sigdelta1}
&&\frac{d\sigma_s}{d\omega d\bm\Delta_\perp}=
\frac{8\alpha\eta^2\varepsilon_q}{\pi\omega\varepsilon_p \Delta^4_\perp}\Phi\,,\nonumber\\
&&\frac{d\sigma_a}{d\omega d\bm\Delta_\perp}=\frac{\pi\mbox{Re}g(\eta)(\varepsilon_p+\varepsilon_q)\Delta}{4\varepsilon_p\varepsilon_q}\frac{d\sigma_s}{d\omega d\bm\Delta_\perp}\,.
\end{eqnarray}
In Fig.~\ref{fig:siga} we show the dependence of $A=\sigma_0^{-1}d\sigma_a/d\omega d\bm\Delta_\perp$ on $\zeta=\Delta_\perp/2m$ for a few values of $t=\varepsilon_q/\varepsilon_p$;
$\sigma_0=\alpha\eta^2\mbox{Re}g(\eta)/(2m^2\omega\varepsilon_p\Delta_\perp)$. This figure confirms our statement that the main contribution to the antisymmetric part of the cross section is given by the region  $\Delta\sim m$.
 Performing integration over $\bm\Delta_\perp$  in \eqref{sigdelta}, we obtain the antisymmetric correction to the spectrum
\begin{eqnarray}\label{sp}
\frac{d\sigma_a}{d\omega}=\frac{\alpha\pi^3\eta^2\mbox{Re}g(\eta)}{4m\omega\varepsilon_p^2}\left[2\frac{\varepsilon_p^2+\varepsilon_q^2}{\varepsilon_p\varepsilon_q}-1\right](\varepsilon_p+\varepsilon_q)\,.
\end{eqnarray}
This result coincides  with  the corresponding result of \cite{LMS2004, LMSS2005}.

\begin{figure}[h]
\centering
\includegraphics[width=0.6\linewidth]{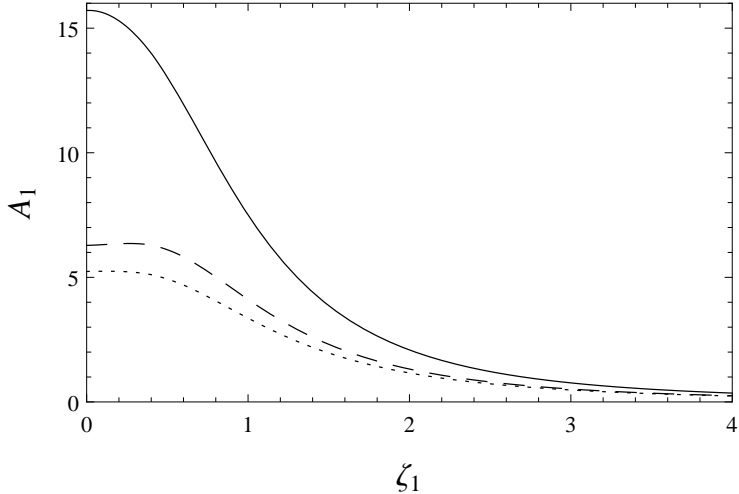}
\setlength{\unitlength}{0.7\linewidth}
\caption{Dependence of $A_1=\sigma_1^{-1}d\sigma_a/d\omega d\bm k_\perp$ on $\zeta_1=k_\perp/m$, for a few values of $t=\varepsilon_q/\varepsilon_p$;
$\sigma_1=\alpha\eta^2\mbox{Re}g(\eta)/(2m^3\omega\varepsilon_p)$: $t=0.25$ (solid curve), $t=0.5$ (dashed curve), and $t=0.75$ (dotted curve).}
\label{fig:siga1}
\end{figure}

In Fig.~\ref{fig:siga1} we show the dependence of $A_1=\sigma_1^{-1}d\sigma_a/d\omega d\bm k_\perp$ on  $\zeta_1=k_\perp/m$ for a few values of $t=\varepsilon_q/\varepsilon_p$;
$\sigma_1=\alpha\eta^2\mbox{Re}g(\eta)/(2m^3\omega\varepsilon_p)$. Here $\bm k_\perp$ is the component of $\bm k$ perpendicular to the vector $\bm p$, $\bm k_\perp= -\omega \bm p_\perp/p$.
The result is obtained by numerical integration of the differential cross section $d\sigma_a(\bm p,\bm q,\bm k,\eta)$ over $d\bm q_\perp$.
As is known (see, e.g., \cite{BLP1982}), the cross section $d\sigma_{brem}(\omega,\varepsilon_p,k_\perp)/d\omega d\bm k_\perp$ of bremsstrahlung ($\varepsilon_p$ is the initial electron energy) can be obtained from the cross section $d\sigma_{photo}(\omega,\epsilon_p, p_\perp)/d\varepsilon_p d\bm p_\perp$ of  photoproduction  ($\varepsilon_p$ is the positron energy)  by the relation
\begin{eqnarray}\label{rel}
\frac{d\sigma_{brem}(\omega,\epsilon_p, k_\perp)}{d\omega d\bm k_\perp}=\frac{\varepsilon_p^2 d\sigma_{photo}(-\omega,-\epsilon_p, p_\perp)}{\omega^2d\varepsilon_p d\bm p_\perp}\,,
\end{eqnarray}
where $p_\perp=p k_\perp/\omega$. The antisymmetric part of $d\sigma_{photo}(\omega,\epsilon_p, p_\perp)/d\varepsilon_p d\bm p_\perp$ was obtained in Ref.~\cite{LMS2012}. Using Fig. 2 in that paper and Eq.~\eqref{rel}, we find that our result shown in Fig.~\ref{fig:siga1} is in agreement with the corresponding result in Ref.~\cite{LMS2012}.

\section{Charge asymmetry in  high-energy  bremsstrahlung from muons}\label{CAM}
The charge asymmetry in the  differential cross section of  high-energy  $\mu^+\mu^-$ photoproduction  in the electric field of a heavy atom was investigated in detail in Ref.~\cite{DLMR2014}.
The deviation of the nuclear electric field from the Coulomb field at small distances due to the finite nuclear radius $R$ (nuclear size effect) is crucially important for the charge asymmetry.  Though the Coulomb corrections to the total cross section are negligibly small, it was shown in Ref.~\cite{DLMR2014} that the charge asymmetry is  not negligible  for selected final states of   $\mu^+$ and $\mu^-$. In this section we study the charge asymmetry in the  differential cross section of  high-energy   bremsstrahlung  from muons in the  field of a heavy atom.

\begin{figure}[h]
\centering
\includegraphics[width=0.6\linewidth]{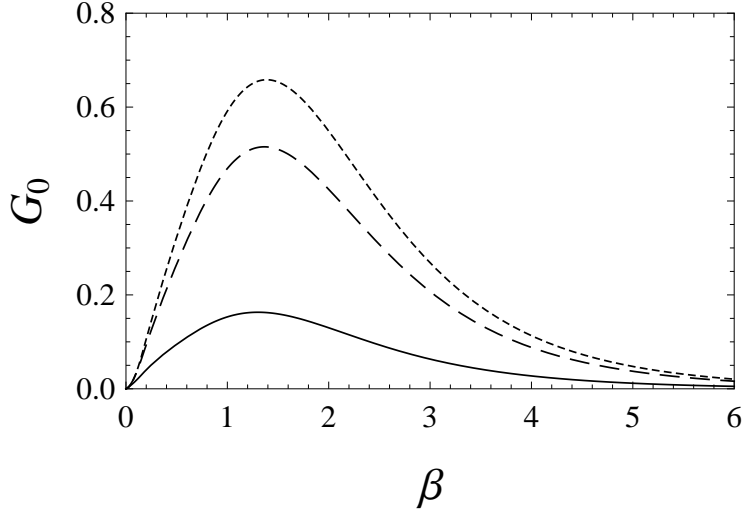}
\setlength{\unitlength}{0.7\linewidth}
\caption{Dependence of $G_0=|N_{0}|^2/|N_{0B}|^2-1$ on $\beta=\Delta/\Lambda$ at $\beta\gg 1/(r_{scr}\Lambda)$ and a few values of $\eta$;
$\eta=0.34$ (Ag, solid curve), $\eta=0.6$ (Pb, dashed curve), and $\eta=0.67$ (U, dotted curve).}
\label{fig:G0}
\end{figure}

 We write the Fourier transformation $V_F(\Delta^2)$ of the potential $V(r)$ as
 \begin{equation}
V_F(\Delta^2)=-\frac{4\pi\eta{ F}(\Delta^2)}{\Delta^2}\,,
\end{equation}
 where  ${F}(\Delta^2)$ is the form factor which differs essentially  from unity at $\Delta\gtrsim 1/R$ and $Q\lesssim 1/r_{scr}$. Let us first discuss the Coulomb corrections to the symmetric part of the cross section calculated in the leading quasiclassical approximation. In this case the cross section $d\sigma_s$ depends on the parameters of the field via the factor $N_0$ \eqref{Mres}. In the Born approximation
\begin{eqnarray}\label{N0B}
&&N_{0B}=\frac{2i}{\omega m^2\Delta^2}\int d\bm r \exp(-i\bm\Delta\cdot\bm r)\bm\Delta\cdot\bm\nabla_\perp V(\bm r)=-\frac{2V_F(\Delta^2)}{\omega m^2}\,.
\end{eqnarray}
We define the Coulomb corrections ${\cal N}_{0}$ to the quantity $|N_{0}|^2$ as
\begin{eqnarray}\label{N02C}
&&{\cal N}_{0}=|N_{0}|^2-|N_{0B}|^2\,.
\end{eqnarray}
The quantity  ${\cal N}_{0}$ vanishes  at $r_{scr}^{-1}\ll \Delta\ll R^{-1}$ and  has two peaks:  at $\Delta\sim r_{scr}^{-1}$ and  at  $\Delta\sim R^{-1}$. The contributions of these peaks to the integral $\int \Delta^2{\cal N}_{0} d\bm\Delta_\perp$ are \cite{LMSS2005}
\begin{eqnarray}\label{N02Cint}
&&\int \Delta^2{\cal N}_{0} d\bm\Delta_\perp=  \mp \frac{128\pi^3\eta^2 f(\eta)}{\omega^2m^4}\,, \quad f(\eta)=\mbox{Re}\psi(1+i\eta)-\psi(1)\,,
\end{eqnarray}
where $\psi(x)=d\ln\Gamma(x)/dx$, the negative contribution corresponds to the  peak at  $\Delta\sim r_{scr}^{-1}$, and the positive contribution corresponds to the  peak at  $\Delta\sim R^{-1}$. These two contributions are the universal functions of $\eta$ independent of the form of the potential in the regions $r\sim r_{scr}$ and $r\sim R$, while the function
${\cal N}_{0}$ is very sensitive to  the form of the potential in these regions \cite{OlsenMW1957,LMSS2005}. Since  $m\ll R^{-1}$ for electrons,  only the region  $r\sim r_{scr}$ gives the Coulomb corrections to $d\sigma_s/d\omega$ \cite{OlsenMW1957},
\begin{eqnarray}\label{eq:spectrphot}
&&\frac{d\sigma_{C}}{d\omega}=-\frac{4\alpha\eta^2f(\eta)}{m^2\omega}\left(t^2-\frac23 t+1\right)\,,\quad t=\frac{\varepsilon_q}{\varepsilon_p}\,.
\end{eqnarray}
For muons  $m_{\mu}\gg R^{-1}$, so that the sum of the contributions from  both peaks,   $\Delta\gtrsim 1/R$ and $Q\lesssim 1/r_{scr}$, gives the total Coulomb corrections to $d\sigma_s/d\omega$. As a result, the total Coulomb corrections vanish, see Eq.\eqref{N02Cint}. However, the Coulomb corrections to the differential cross section at $\Delta\sim R^{-1}$ are large. To illustrate this statement, we consider the form factor $F(\Delta^2)$ in the form
\begin{equation}\label{FF}
 F(\Delta^2)=\frac{\Lambda^2}{\Delta^2+\Lambda^2}\,,
\end{equation}
where $\Lambda\sim 60\,\mbox{MeV}$ for heavy nuclei. This form is valid for $\Delta\gg r_{scr}^{-1}$, where the factor $N_0$ is given by
\begin{eqnarray}\label{N0}
&&N_0=\frac{8\pi\eta}{\omega m^2\Delta^2}\int_0^\infty d\rho\,J_1(\rho)\left[1-\frac{\rho}{\beta}K_1\left(\frac{\rho}{\beta}\right)\right]
\exp\Bigg\{-2i\eta\Bigg[\ln\frac{\rho}{2}+K_0\left(\frac{\rho}{\beta}\right)\Bigg]\Bigg\}\,,\nonumber\\
&&N_{0B}=\frac{8\pi\eta}{\omega m^2\Delta^2(1+\beta^2)}\,,\quad \beta=\frac{\Delta}{\Lambda}\,.
\end{eqnarray}
Here $J_n(x)$ is the Bessel function and $K_n(x)$ is the  modified Bessel function of the second kind.
In Fig.~\ref{fig:G0}  we show the dependence of $G_0=|N_{0}|^2/|N_{0B}|^2-1$ on $\beta=\Delta/\Lambda$ for a few values of $\eta$.
Note that very narrow peak at $\Delta\sim r_{scr}^{-1}$ ($\delta\beta\sim 1/(r_{scr}\Lambda)\ll 1$)  is not shown in this figure. The dependence of the peak on the shape of the atomic potential at $\Delta\sim r_{scr}^{-1}$ was investigated in detail in Ref.~\cite{LMSS2005}.   It is seen from Fig.~\ref{fig:G0}  that the Coulomb corrections to $|N_{0}|^2$ are significant in the region $\Delta/\Lambda\sim 1$.

\begin{figure}[H]
\centering
\includegraphics[width=0.6\linewidth]{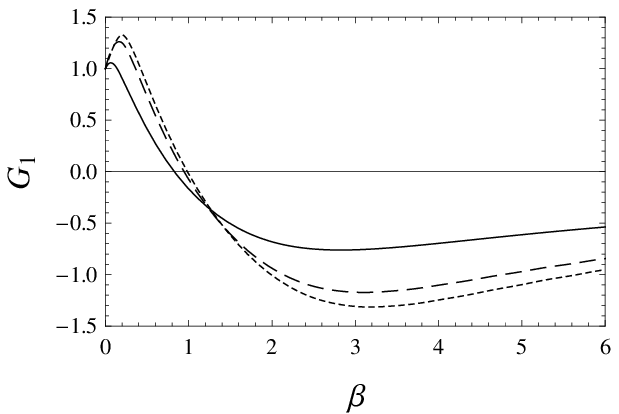}
\setlength{\unitlength}{0.7\linewidth}
\caption{Dependence of  $G_1=\Sigma_R^{-1}\mbox{Re}N_0N_1^*/|N_{0}|^2$ on $\beta=\Delta/\Lambda$ \eqref{G1G2}, for a few values of $\eta$;
$\eta=0.34$ (Ag, solid curve), $\eta=0.6$ (Pb, dashed curve), and $\eta=0.67$ (U, dotted curve).}
\label{fig:G1}
\end{figure}

\begin{figure}[H]
\centering
\includegraphics[width=0.6\linewidth]{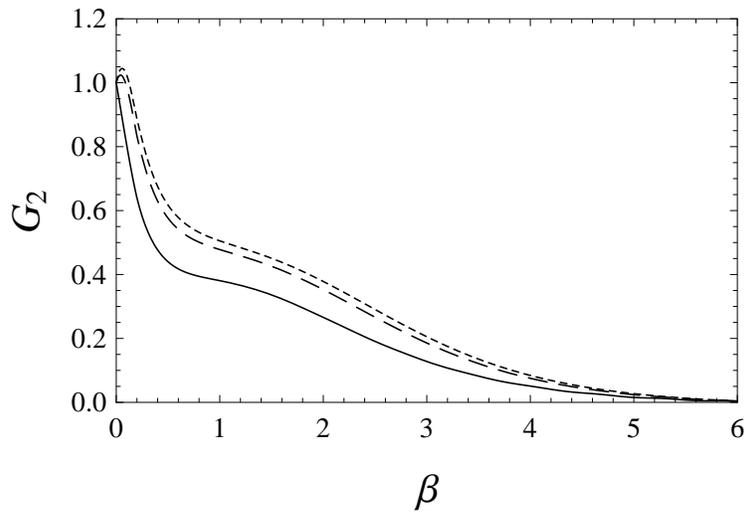}
\setlength{\unitlength}{0.7\linewidth}
\caption{Dependence of $G_2=\Sigma_I^{-1}\mbox{Im}N_0N_1^*/|N_{0}|^2$ on $\beta=\Delta/\Lambda$ \eqref{G1G2}, for a few values of $\eta$;
$\eta=0.34$ (Ag, solid curve), $\eta=0.6$ (Pb, dashed curve), and $\eta=0.67$ (U, dotted curve).}
\label{fig:G2}
\end{figure}

Let us consider the factor $N_1$ \eqref{Mres}. In the lowest in $\eta$ approximation it reads
\begin{eqnarray}\label{N1B}
&&N_{1B}=\frac{{\cal J}(\Delta)}{\omega m^2\varepsilon_p\varepsilon_q}\,,\nonumber\\
&&{\cal J}(\Delta)=\int\frac{d\bm s}{(2\pi)^3}[V_F(Q_+)V_F(Q_-)+
(\Delta^2-4s_\parallel^2)V_F(Q_+)V_F'(Q_-)]\,,\nonumber\\
&&Q_\pm=(\bm s\pm \bm \Delta/2)^2\,,\quad s_\parallel=\bm s\cdot\bm\Delta/\Delta\,,
\end{eqnarray}
where $V_F'(Q)=\partial V_F(Q)/\partial Q$, see Ref.\cite{DLMR2014}. For the form factor \eqref{FF}, the explicit form of ${\cal J}(\Delta)$  is given in Ref. \cite{DLMR2014}:
\begin{eqnarray}\label{calJ}
&&{\cal J}(\Delta)=\frac{2\pi^2\eta^2}{\Delta}{\cal F}(\beta)\,,\quad \beta=\frac{\Delta}{\Lambda}\,,\nonumber\\
&&{\cal F}(\beta)=1+\frac{2}{\pi}\arcsin\frac{\beta}{\sqrt{\beta^2+4}}-\frac{4}{\pi}\arcsin\frac{\beta}{\sqrt{\beta^2+1}}-\frac{12 \beta}{\pi (\beta^2+4)(\beta^2+1)}\,.
\end{eqnarray}
The exact in $\eta$ factor $N_1$ at  $\Delta\gg r_{scr}^{-1}$ is given by
\begin{eqnarray}\label{N1exact}
&&N_{1}=\frac{2\pi^2\eta^2}{\omega m^2\varepsilon_p\varepsilon_q\Delta}\int_0^\infty\!\!\!\int_0^\infty dx d\rho \,{\cal F}(\beta x/\rho)
\,J_0(\rho)J_0(x)\nonumber\\
&&\times\exp\Bigg\{-2i\eta\Bigg[\ln\frac{\rho}{2}+K_0\left(\frac{\rho}{\beta}\right)\Bigg]\Bigg\}\,.
\end{eqnarray}
To demonstrate the influence of the nuclear size effect on the ratios  $S_1/S_0$ and  $S_2/S_0$ \eqref{M2}, we plot in Figs. \ref{fig:G1} and \ref{fig:G2} the quantities $G_{1}$ and $G_{2}$,
\begin{eqnarray}\label{G1G2}
&&G_{1}=\frac{\mbox{Re}N_0N_1^*}{|N_0|^2\Sigma_R}\,,\quad \Sigma_R=\frac{\pi\mbox{Re}g(\eta)\Delta}{4\varepsilon_p\varepsilon_q}\,,\nonumber\\
&&G_{2}=\frac{\mbox{Im}N_0N_1^*}{|N_0|^2\Sigma_I}\,,\quad \Sigma_I=\frac{\pi\mbox{Im}g(\eta)\Delta}{4\varepsilon_p\varepsilon_q}\,,
\end{eqnarray}
as a function of $\beta=\Delta/\Lambda$. For a pure Coulomb field $G_{1}=G_{2}=1$ \eqref{M2C}.

It is seen that the quantities $G_{1}$ and $G_{2}$  decrease rapidly with increasing $\beta$ at $\beta\lesssim 1$.

\section{Conclusion}\label{concl}
We have investigated in detail  the charge asymmetry  in the process  of  high-energy
 bremsstrahlung  in the  field of a heavy atom. The charge asymmetry arises due to the  account for the first quasiclassical correction to the differential cross section of the process.
 The results  are  exact in the parameters of the atomic field and are  valid even for $\eta\sim 1$, they   take  into account the effect of screening  and the nuclear size effect. The latter is important for high-energy  bremsstrahlung from muons where the charge asymmetry is very sensitive to the shape of the nuclear form factor. It is shown that the Coulomb corrections essentially modify the  charge asymmetry  as compared with the leading in $\eta$ result already for the relatively small $\eta$. In the experimental region of interest, where $\varepsilon_p\gg p_\perp\gg m$ and $\varepsilon_q\gg q_\perp\gg m$ but $\Delta\lesssim 1/R$, the asymmetry  can be as large as  a few tens of percent. For the longitudinal polarization of the initial charged particle, due to the account for the first quasiclassical correction, the differential cross section reveals  the  asymmetry   with respect to the replacement $\varphi\rightarrow-\varphi$, where $\varphi$ is the azimuth angle between the photon momentum $\bm k$ and the momentum $\bm q$ of the final charged particle  in the frame where the z axis is directed along $\bm p$. Due to account for the first quasiclassical correction, our results for the differential cross section of  high-energy  bremsstrahlung have essentially higher precision  than the famous results in Ref.~\cite{OlsenMW1957} and should be taken into account in  precision experiments and at data analysis in  detectors.

\section*{Acknowledgement}
We are grateful to R.N.~Lee for important discussions. This work has been  supported by Russian Science
Foundation (project N 14-50-00080). Investigation of the charge asymmetry from muons has been also supported  in part by the RFBR (Grant No. 14-02-00016).

\section*{Appendix}
In this appendix, following the method of \cite{LMSS2005}, we derive the expression \eqref{ALLA} for the  quantity $A_{00}$,
\begin{eqnarray}\label{fgapp}
&&A_{00}=\int \!\!d\bm r\,\exp{(-i\bm k\cdot\bm r )}f_0g_0\,,
\end{eqnarray}
where the function $f_0(\bm r,\bm q)$ is given in  \eqref{f0} and $g_0(\bm r,\bm p)=f_0(\bm r,-\bm p)$.   Other  quantities in \eqref{ALLA} are calculated in the same way.
We split the integration region into two,  $z>0$ and $z<0$, and denote the corresponding contributions to $A_{00}$   as $A_{00}^+$ and $A_{00}^-$. For $z>0$, the function $f_0$ has a simple  eikonal form
\begin{eqnarray}\label{f0eik}
f_0(\bm r,\bm q)=e^{-i\bm q\cdot\bm r}\exp\left[-i\int_0^\infty dx V(\bm r+x\bm n_{\bm q}) \right]\,,
\end{eqnarray}
so that
\begin{eqnarray}\label{Ap}
A^+_{00}&=&\int_{z>0} d \bm r\int \frac{d\bm Q}{i \pi} \exp\left\{ iQ^2 -i\bm\Delta\cdot\bm r -i\int_0^\infty dx [V(\bm r_x)+V(\bm r+x\bm n_{\bm q})]\right\}\nonumber \\
&&\times\left[1+\frac{i}{2\varepsilon_p}\int_0^\infty dx\int_0^x dy(x-y)\bm\nabla_{\perp}V(\bm r_x)\cdot\bm\nabla_{\perp}V(\bm r_y) \right]\,,
\end{eqnarray}
where $\bm{r_x}=\bm r-x\bm n_{\bm p}+\bm Q\sqrt{2r/p}$.
Within our accuracy we can  replace   the quantity $V(\bm r+x\bm n_{\bm q})$ in \eqref{Ap} by $V(\bm r+x\bm n_{\bm q}+\bm Q\sqrt{2r/p})$,   shift $\bm\rho\to\bm\rho-\bm Q\sqrt{2r/p}$, and take the integral over $\bm Q$. We obtain
\begin{eqnarray}\label{Ap1}
A^+_{00}&=&\int_{z>0} d \bm{r} \exp\left\{ -i\frac{z}{2p}\Delta_{\perp}^2 -i\bm\Delta\cdot\bm r-i\int_0^\infty dx [V(\bm r-x\bm n_{\bm p})+V(\bm r+x\bm n_{\bm q})]\right\}\nonumber \\ &&\times\left[1+\frac{i}{2\varepsilon_p}\int_0^\infty dx\int_0^x dy(x-y)\bm\nabla_{\perp}V(\bm r-x\bm n_{\bm p})\cdot\bm\nabla_{\perp}V(\bm r-y\bm n_{\bm p}) \right]\,.
\end{eqnarray}
In the same way, we obtain
\begin{eqnarray}\label{Am1}
A^-_{00}&=&\int_{z<0} d \bm{r} \exp\{ i\frac{z}{2q}\Delta_{\bot}^2 -i\bm\Delta\cdot\bm r-i\int_0^\infty dx [V(\bm r-x\bm n_{\bm p})+V(\bm r+x\bm n_{\bm q})]\}\nonumber \\ &&\times\left[1+\frac{i}{2\varepsilon_q}\int_0^\infty dx\int_0^x dy(x-y)\bm\nabla_{\perp}V(\bm r+x\bm n_{\bm q})\cdot\bm\nabla_{\perp}V(\bm r+y\bm n_{\bm q}) \right]\,.
\end{eqnarray}
There are two overlapping regions of the momentum transfer $\Delta$:
\begin{eqnarray}\label{regions}
&&\mbox{I}.\, \Delta\ll\frac{m \omega}{\varepsilon_p}\nonumber\\
&&\mbox{II}.\, \Delta\gg\Delta_{min}=\frac{m^2\omega}{2\varepsilon_p\varepsilon_q}\,.
\end{eqnarray}
In the first region, one can neglect the term proportional to $\Delta_{\bot}^2$ in the exponents in  \eqref{Ap1} and \eqref{Am1}. Then the sum $A_{00}=A_{00}^++A_{00}^-$ reads
\begin{eqnarray}\label{1reg}
&&A_{00}=\int d \bm r  \exp \left\{-i\bm\Delta\cdot\bm r-i\int_0^\infty dx [V(\bm r-x\bm n_{\bm p})+V(\bm r+x\bm n_{\bm q})\right\}\nonumber\\
&&\times\Bigg[1+\frac{i}{2\varepsilon_p}\int_0^\infty dx\int_0^x dy(x-y)\bm\nabla_{\perp}V(\bm r-x\bm n_{\bm p})\cdot\bm\nabla_{\perp}V(\bm r-y\bm n_{\bm p})\nonumber\\
&&+\frac{i}{2\varepsilon_q}\int_0^\infty dx\int_0^x dy(x-y)\bm\nabla_{\perp}V(\bm r+x\bm n_{\bm q})\cdot\bm\nabla_{\perp}V(\bm r+y\bm n_{\bm q})\Bigg]\,.
\end{eqnarray}
In the prefactor we  make the replacement $\bm n_{\bm p},\bm n_{\bm q}\rightarrow\bm \nu$, and in the exponent we  take into account the linear term of  expansion  in
 $\bm n_{\bm p}-\bm\nu$ and  $\bm n_{\bm q}-\bm\nu$ of the integral. Besides, in the arguments of the functions $V(\bm r+y\bm\nu)$ and $V(\bm r-y\bm\nu)$ we make the substitutions $z\rightarrow z-y$  and  $z\rightarrow z+y$, respectively. After that we take the integral over $y$  and obtain  the contribution of the first region
\begin{eqnarray}\label{1reg}
&&A_{00}=\int d \bm r  \exp[-i\bm\Delta\cdot\bm{r}-i\chi(\rho)]\bm \nabla_{\perp}V(\bm r)\nonumber\\
&&\cdot\Bigg[i\frac{\bm \theta_{qp}}{\Delta_z^2}-\frac{\omega}{2\Delta_z\varepsilon_p\varepsilon_q}\int_0^\infty dx x\bm\nabla_{\perp}V(\bm r-x\bm\nu)\Bigg]\,,\nonumber\\
&&\chi(\rho)=\int_{-\infty}^\infty V(z,\bm\rho)dz\,,
\end{eqnarray}
where $\Delta_z=\bm\nu\cdot\bm\Delta$ and $\bm \theta_{qp}=\bm q_\perp/q-\bm p_\perp/p$.

Now we pass to the calculation of $A_{00}$ in the second region \eqref{regions}. In Eq.(\ref{Ap1}) for $A_{00}^+$, we make the replacement $\bm n_q\rightarrow \bm n_p$ and $z\Delta_{\perp}^2/2p\rightarrow (\bm n_q\cdot\bm n)\Delta_{\perp}^2/2p$. The polar angle of $\bm n$ is small, and we can integrate in \eqref{Ap1} over the region  $\bm n_q\cdot\bm n>0$. After the integration over $z$, we obtain
\begin{eqnarray}\label{2regp}
&&A_{00}^+=\frac{1}{\bm\Delta\cdot\bm n_p+\Delta_{\perp}^2/2p}\int d \bm \rho  \exp [-i\bm\Delta_{\perp}\cdot\bm \rho-i\chi(\rho)]\nonumber\\
&&\times\Bigg[-i+\int_{-\infty}^{\infty} dz\int_0^{\infty} dx x\bm\nabla_{\perp}V(\bm r)\cdot\bm\nabla_{\perp}V(\bm r-x\bm n_p) \Bigg]\,.
\end{eqnarray}
The calculation of $A_{00}^-$ is performed quite similarly. As a result we have
\begin{eqnarray}\label{2regm}
&&A_{00}^-=\frac{1}{-\bm\Delta\cdot\bm n_q+\Delta_{\perp}^2/2q}\int d \bm \rho  \exp [-i\bm\Delta_{\perp}\cdot\bm \rho-i\chi(\rho)]\nonumber\\
&&\times\Bigg[-i+\int_{-\infty}^{\infty} dz\int_0^{\infty} dx x\bm\nabla_{\perp}V(\bm r)\cdot\bm\nabla_{\perp}V(\bm r-x\bm n_q) \Bigg]\,.
\end{eqnarray}
Taking into account that
\begin{eqnarray}
&&\bm\Delta\cdot\bm n_p+\Delta_{\perp}^2/2p=-\frac{m^2\omega}{2\varepsilon_p\varepsilon_q \xi_q}\,,\quad -\bm\Delta\cdot\bm n_q+\Delta_{\perp}^2/2q=\frac{m^2\omega}{2\varepsilon_p\varepsilon_q \xi_p},\nonumber\\
&&\xi_p=\frac{m^2}{m^2+\bm{p}_{\perp}^2}\,,\quad \xi_q=\frac{m^2}{m^2+\bm{q}_{\perp}^2}\,,\nonumber
\end{eqnarray}
we obtain for $A_{00}=A_{00}^++A_{00}^- $ in the second region
\begin{eqnarray}\label{sh}
&&A_{00}=\frac{1}{m^4 \omega}\int d \bm\rho  \exp [-i\bm\Delta_\perp\cdot\bm\rho-i\chi(\rho)]
\Bigg[2i\varepsilon_p\varepsilon_q\xi_p\xi_q(\bm p_{\perp}+\bm q_{\perp})\cdot\bm \nabla_{\perp}\chi(\rho) \nonumber\\
&&+m^2(\varepsilon_p\xi_p-\varepsilon_q\xi_q)\int_{-\infty}^{\infty}dz\int_0^{\infty}dx x \bm\nabla_{\perp}V(\bm r-x\bm\nu)\cdot\bm\nabla_{\perp}V(\bm r)\Bigg]\,.
\end{eqnarray}
Now we can compare  \eqref{sh} and \eqref{1reg} and write the expression for $A_{00}$ which is valid in all region of $\Delta$. We finally arrive at the following result
\begin{eqnarray}\label{finalA00}
&&A_{00}=\frac{1}{m^4 \omega}\int d \bm r \exp [-i\bm\Delta\cdot\bm r-i\chi(\rho)]
\Bigg[2i\varepsilon_p\varepsilon_q\xi_p\xi_q(\bm p_{\perp}+\bm q_{\perp}) \nonumber\\
&&+m^2(\varepsilon_p\xi_p-\varepsilon_q\xi_q)\int_0^{\infty}dx x \bm\nabla_{\perp}V(\bm r-x\bm\nu)\Bigg]\cdot\bm\nabla_{\perp}V(\bm r)\,.
\end{eqnarray}

 \end{document}